\documentclass[twocolumn,showpacs,superscriptaddress,preprintnumbers,amsmath,amssymb]{revtex4}

\usepackage{graphicx}
\usepackage{dcolumn}
\usepackage{bm}

\begin{document}

\preprint{APS/123-QED}

\title{Resonances in $^{19}$Ne with relevance to the astrophysically important $^{18}$F(p,$\mathbf{\alpha}$)$^{15}$O reaction}

\author{D.J. Mountford
 \footnote{corresponding author: d.j.mountford@sms.ed.ac.uk}}
\affiliation{%
SUPA,  School of Physics and Astronomy, University of Edinburgh, EH9 3JZ, UK
}%

\author{A.St\,J. Murphy}

\affiliation{%
SUPA,  School of Physics and Astronomy, University of Edinburgh, EH9 3JZ, UK
}%

\author{N.L. Achouri}
\affiliation{%
Laboratoire de Physique Corpusculaire ENSICAEN, CNRS-IN2P3 UMR 6534 et Universite ́ de Caen, F-14050 Caen, France
}%

\author{C. Angulo}
\affiliation{%
Tractebel Engineering, Avenue Ariane 7, B-1200 Brussels, Belgium
}%

\author{J.R. Brown}
\affiliation{
Department of Physics, University of York, YO10 5DD, UK
}%

\author{T. Davinson}
\affiliation{%
SUPA,  School of Physics and Astronomy, University of Edinburgh, EH9 3JZ, UK
}%

\author{F. de Oliveira Santos}
\affiliation{%
Grand Acc$\acute{e}$l$\acute{e}$rateur National d'Ions Lourds (GANIL), CEA/DSM-CNRS/IN2P3, Caen, France
}%

\author{N. de S\'er\'eville}
\affiliation{%
Institut de Physique Nucl$\acute{e}$aire UMR 8608, CNRS-IN2P3/Universit$\acute{e}$ Paris-Sud, F-91406 Orsay, France
}%

\author{P. Descouvemont}
\affiliation{%
Physique Nucl\'eaire Th\'eorique et Physique Math\'ematique, C.P. 229,\\
Universit\'e Libre de Bruxelles (ULB), B 1050 Brussels, Belgium
}%

\author{O. Kamalou}
\affiliation{%
Grand Acc$\acute{e}$l$\acute{e}$rateur National d'Ions Lourds (GANIL), CEA/DSM-CNRS/IN2P3, Caen, France
}%

\author{A.M. Laird}
\affiliation{
Department of Physics, University of York, YO10 5DD, UK
}%

\author{S.T. Pittman}
\affiliation{
Department of Physics and Astronomy, University of Tennessee, Knoxville, Tennessee 37996, USA
}%

\author{P. Ujic}
\affiliation{%
Grand Acc$\acute{e}$l$\acute{e}$rateur National d'Ions Lourds (GANIL), CEA/DSM-CNRS/IN2P3, Caen, France
}%

\author{P.J. Woods}
\affiliation{%
SUPA,  School of Physics and Astronomy, University of Edinburgh, EH9 3JZ, UK
}%

\date{\today}

\begin{abstract}
\noindent
The most intense gamma-ray line observable from novae is likely to be from positron annihilation associated with the decay of $^{18}$F. 
The uncertainty in the destruction rate of this nucleus through the $^{18}$F(p,$\alpha$)$^{15}$O reaction 
presents a limit to interpretation of any future observed gamma-ray flux. Direct measurements of the cross section of both this reaction and the 
$^{18}$F(p,p)$^{18}$F reaction have been performed between center of mass energies of 0.5 and 1.9~MeV.
Simultaneous fits to both data sets with the R-Matrix formalism reveal several resonances, 
with the inferred parameters of populated states in $^{19}$Ne 
in general agreement with previous measurements. Of particular interest, extra strength has been observed
above~$E_{CM}\sim$1.3~MeV in the $^{18}$F(p,p)$^{18}$F reaction and between 1.3--1.7~MeV in the 
$^{18}$F(p,$\alpha$)$^{15}$O reaction. This is well described by a broad 1/2$^{+}$ state, consistent with both a recent 
theoretical prediction and an inelastic scattering measurement. The astrophysical implications of a broad sub-threshold
partner to this state are discussed.
\end{abstract}

\maketitle


\section{\label{sec:Intro}Introduction}

\noindent 
Novae are the most common astrophysical explosion sites in the Universe and provide an excellent opportunity 
for the study of the nucleosynthesis of radioisotopes through detection of their gamma-ray emission. 
The current INTEGRAL satellite includes such an objective within its mission goals. A robust prediction
of simulations of CO- and ONe-type novae events is that their gamma-ray emission will be
dominated by the 511~keV gamma-rays produced by positron annihilation following the $\beta^{+}$ decay of 
$^{18}$F~\cite{Hernanz1999}. Key reasons for the high flux are that this isotope is produced relatively abundantly, 
and its lifetime of $\sim$158~minutes is well matched to the timescale for nova ejecta to become transparent to 
gamma-ray emission. There remains however a large uncertainty on the absolute flux, and by
implication the detectability distance for novae, because of uncertainty in the rates of the nuclear 
reactions producing and destroying $^{18}$F. Of these, the reaction which contributes
the most uncertainty to the final abundance of $^{18}$F is that of $^{18}$F(p,$\alpha$)$^{15}$O.
It is this reaction which is the focus of the present work.

Despite significant recent effort expended on determination of this reaction rate, the situation 
remains unclear. In the energy region of interest, several states in $^{19}$Ne may make resonant contributions
to the $^{18}$F(p,$\alpha$)$^{15}$O reaction rate (see~\cite{Nesaraja2007} for a recent summary). 
$S$- and $p$-wave resonances just above and below the $^{18}$F+p threshold at 6411~keV have proven 
difficult to measure, although some recent progress has been made~\cite{Adekola2011}. The impact of interference between 
states of the same spin-parity is still unclear, although again, some recent progress has been made~\cite{Beer2011}.
It is also expected that there are additional resonances in the region of interest that are yet to be
seen. One such prediction is for a broad $J^{\pi}=1/2^{+}$ state, below threshold, but which has sufficient
width to play an important role in the nova burning energy range~\cite{Dufour2007}.  The Generator Coordinate
Method model that made this prediction also suggests the existence of another broad 1/2$^{+}$ state 
($\Gamma_{p} =$ 157~keV, $\Gamma_{\alpha} =$ 139~keV) at a resonance
energy around 1.49~MeV (E$_{x}$($^{19}$Ne) $=$ 7894~keV). Data from a recent $^{1}$H($^{19}$Ne,p)$^{19}$Ne$^{*}$(p)$^{18}$F 
inelastic scattering measurement revealed a broad feature beneath several well defined
peaks that is consistent with the presence of such a state~\cite{Dalouzy2009}. However, a direct measurement 
using the thick target method~\cite{Murphy2009} revealed no such feature, though a narrow state of different
spin was observed in the vicinity. This measurement only took data up to $E_{CM}\sim1.6~MeV$, 
hence the assignment of parameters to this state is hampered by its vicinity to this high energy cut off.


\section{\label{sec:exp}Experimental Work}

\begin{figure}[htb]
\includegraphics[scale=0.45]{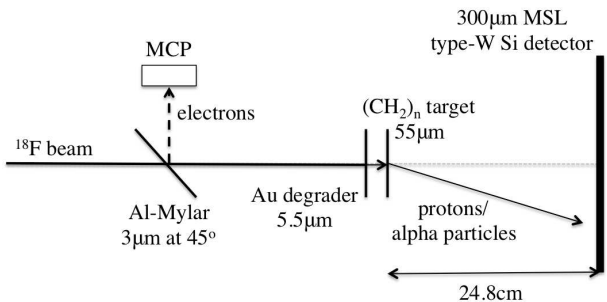}
\caption{\label{fig:setup}Schematic layout of the experimental set up. The $^{18}$F was stopped in a thick CH$_{2}$ target. Recoiling proton and alpha particles were detected in a double sided silicon strip detector.}
\end{figure}

\noindent An experiment was carried out at the GANIL-SPIRAL facility, Caen, in April 2010.
A schematic diagram of the experimental set up used is shown in Figure~\ref{fig:setup}.
A 95~MeV/A primary beam of $^{20}$Ne bombarded a thick carbon target.  
Secondary $^{18}$F ions were extracted in the molecular form HF, ionized in an 
ECR ion source, and post-accelerated with the CIME 
cyclotron~\cite{Duval1996} to form a secondary radioactive ion beam of energy 3.924~MeV/A. 
The typical $^{18}$F intensity was $\sim2\times10^{4}$pps. 
The beam optics were tuned to deliver ions of mass-to-charge
ratio equal to 2, $\it{i.e.}$ a 9$^{+}$ charge state for $^{18}$F ions. This was achieved through use of a 
thin carbon stripper-foil placed in the beam line after the CIME cyclotron
and was motivated by the desire to eliminate expected contamination with $^{18}$O ions. 

\begin{figure}[htb]
\includegraphics[scale=0.35]{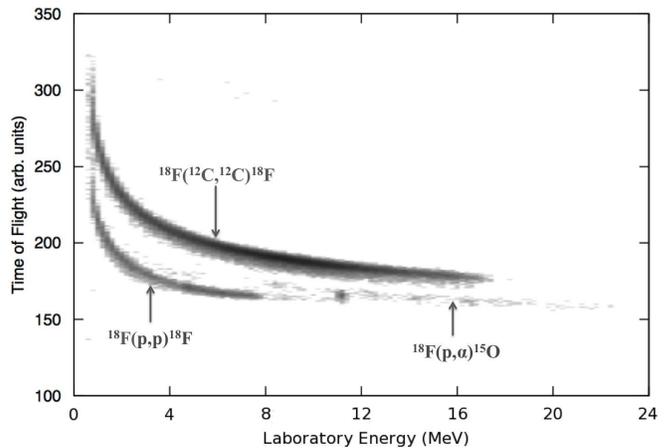}
\caption{\label{fig:ToF}The time difference between beam ions traversing the MCP 
foil and particles being detected in the DSSD, as a function of the detected particle
energy. Three loci are observed: a proton locus (lower left), an alpha particle locus (middle),
and a carbon ion locus from $^{18}$F($^{12}$C,$^{12}$C)$^{18}$F events. 
At $\sim$11~MeV, a $^{4}$He ion beam contaminant is observed.}
\end{figure}

A $5.5\pm0.3~\mu$m Au foil was mounted on
the upstream face of the target ladder, degrading the $^{18}$F beam to an energy of 1.7~MeV/A.
The target then consisted of $55\pm4~\mu$m of low density PTFE (CH$_{2}$ polymer), thick enough to 
stop the beam, but thin enough to allow light ions to escape.
Protons and alpha particles, emitted from $^{18}$F(p,p)$^{18}$F and $^{18}$F(p,$\alpha$)$^{15}$O reactions
in the target, were detected in a 50~mm~$\times$~50~mm 
double sided silicon strip detector (Micron Semiconductors Ltd. type-W~\cite{MSL})
located $248\pm1$~mm downstream of the target. Carbon ions were also observed 
from $^{18}$F($^{12}$C,$^{12}$C)$^{18}$F scattering.
To provide a local reference time for each ion delivered, a 3~$\mu$m aluminised Mylar 
foil at 45$^{\circ}$ to the beam axis and a microchannel plate (MCP) were positioned 
upstream. Employing energy versus time of flight, all three particle 
species were well separated and identified, as shown in Figure~\ref{fig:ToF}. 
The most intense locus in Figure~\ref{fig:ToF} corresponds to carbon ions, 
the next most intense to protons, and the low intensity band between the two
corresponds to alpha-particles. 

Possible contamination of the beam by  $^{18}$Ne ions was 
searched for by inserting a Hamamatsu photodiode detector~\cite{Hamamatsu} 
with a thin aluminum degrader entrance foil into the target position. 
The beam was found to split into two components, with the lower energy component
consistent with $^{18}$Ne at an intensity $\sim3\%$ of that of $^{18}$F, thus 
contributing negligibly to the observed proton yield~\cite{Angulo2003}.
As can be seen in Figure~\ref{fig:ToF}, there appears to be an additional contribution to the 
alpha-particle locus ($E_{lab} \sim$ 11~MeV). 
Gating on these events revealed that they were impacting the center of the detector, {\em i.e.}
these events were aligned with the beam axis and were most likely due to $^{4}$He 
beam contaminant. These events were discounted from further analysis.

For a thick-target experiment, with protons undergoing elastic scattering, 
and for alpha-particles originating from (p,$\alpha$) reactions, the detected 
energy and angle of a particle is uniquely related to the center of mass energy 
of the scattering/reaction. By consideration of all possible target depths at which reactions
might occur and angles to which particles might be detected, an angle-dependent 
algorithm was generated mapping laboratory energy to centre of mass energy.
Energy losses in the degrader, the target and in detector dead layers were based on 
SRIM2008.04~\cite{Ziegler1985}. The algorithm was then applied to each event to generate center of 
mass energy spectra for the $^{18}$F(p,p)$^{18}$F and $^{18}$F(p,$\alpha$)$^{15}$O reactions. 
The resulting spectra are presented in Figure~\ref{fig:Result}.

Contributions to the 
energy resolution of these spectra arise from several factors, including: the
geometric angular resolution of the detector; the intrinsic energy resolution of the detector and associated 
electronics; the energy and angular straggling of the beam and ions in passing through degraders,
target and detector dead layers; uncertainty in the detector alignment with respect to the beam; beam
divergence and beam spot size at the target. A significant additional uncertainty may arise if the absolute energy loss corrections are inaccurate, either because of insufficient knowledge of stopping powers,
or due to incorrect target thickness measurements. The complexity of estimating the overall effect of these
factors, especially given the use of a thick degrader, 
together with the potential sensitivity of the subsequent analysis on correct determination of the energies and
energy resolutions, warranted the development of a dedicated Monte Carlo simulation of the above effects. 
Energy losses, energy straggling and angular straggling were based on SRIM2008.04~\cite{Ziegler1985}, the intrinsic 
energy resolution of the detectors  was 15~keV for protons and 25~keV for alpha particles~\cite{Steinbauer1994}, and
the beam divergence (0.5 degrees) and beam spot size (10 mm) were determined during the experiment.
Target and degrader thicknesses were determined with calibrated alpha-particle energy loss 
measurements and cross-checked by measurements of mass per unit area. 
Both reactions of interest were simulated, with the angular distribution of
reactions assumed to be isotropic in the center of mass. 
Energy resolution for the detected alpha particles (rms, in the lab frame) was found to range between 40 and 70~keV 
for scatters occurring at center of mass energies of 0.6~MeV, increasing to between 
60 and 160~keV for center of mass energy of 1.9~MeV while for the protons, this was between 30 and 40~keV at 0.6~MeV, increasing to between 25 and 50~keV at 1.9~MeV. The larger resolutions correspond to wider angle scatters in both cases.

\section{\label{sec:res}Results and Discussion}

\begin{figure}[htb]
\includegraphics[scale=0.55]{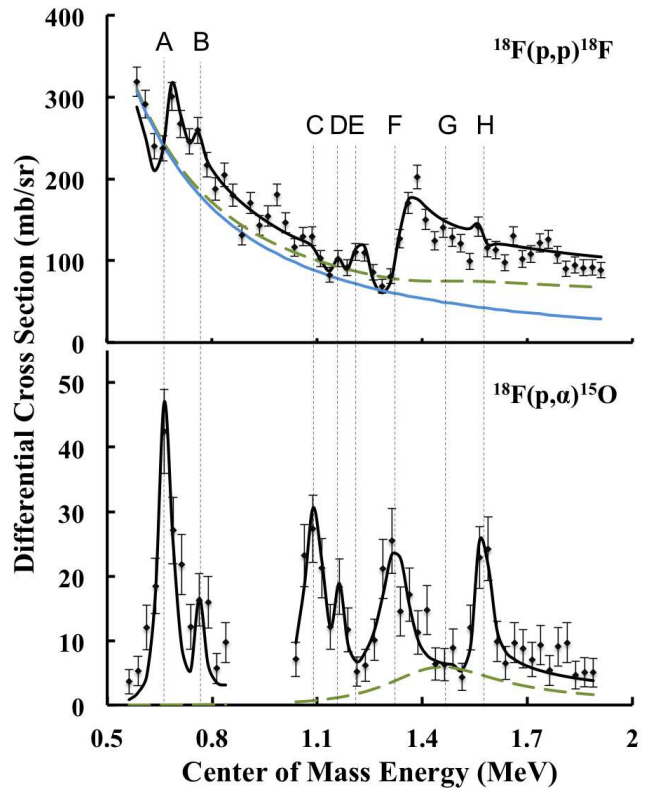}
\caption{\label{fig:Result}(Color online.) Differential cross sections of both $^{18}$F(p,p)$^{18}$F and $^{18}$F(p,$\alpha$)$^{15}$O reactions as a function of center of mass energy. A simultaneous $R$-matrix fit, calculated of a center of mass angle of 176$^{\circ}$, is shown by the solid black line with the 1/2$^{+}$ contribution shown in long-dashed green and (in the elastic scattering case) the Coulomb contribution is in pale blue. The 1/2$^{+}$ contribution is consistent with a predicted broad state~\cite{Dufour2007}. The combined fit has a $\bar{\chi}^{2}$ of 1.633}
\end{figure}

\noindent 
The center of mass excitation functions for the $^{18}$F(p,p)$^{18}$F and 
$^{18}$F(p,$\alpha$)$^{15}$O reactions are shown in Figure~\ref{fig:Result}. 
Data from all detector pixels (all scattering angles) are included as the limited number of events precluded 
projection of angular distributions. Data have been removed near E$_{c.m.}$=0.92~MeV
in the $^{18}$F(p,$\alpha$)$^{15}$O channel because of the contamination due to $^{4}$He ions present in the
beam.
Furthermore, despite the good intrinsic 
energy resolution determined from the Monte Carlo simulations, 
the limited statistics have required that 
the spectra are binned at 25~keV (CM). 
Several resonant structures are
observed of which the most prominent is that at 665~keV, due to the well known 7076~keV $J^{\pi}=3/2^{+}$ 
state in $^{19}$Ne~\cite{Bardayan2001}.  While the relative normalization between the
two data sets was maintained, the absolute normalization was adjusted to provide consistency with the
known differential cross section in the vicinity of this peak; as such no parameters were extracted from 
this data set for this resonance.  It should be noted though that the observed width of this peak
matches well the rather precise widths found in previous studies, providing
some validation of the Monte Carlo used to estimate the energy resolution.

Interpretation of the data, aided by $R$-matrix calculations informed by previous results in the literature, revealed  
6 further resonant structures present in the excitation functions. Their parameters are given in table~\ref{tab:Results}, labelled B-F and H. In addition, there is also additional strength in the region of 1.3--1.7~MeV. 
An additional broad seventh state, here labelled G, has been included to account for this additional strength. 

\begin{table*}
\caption{\label{tab:Results} Tabulation of the resonance parameters extracted from the present data 
when all resonances are allowed to vary within the minimisation procedure. The resulting $R$-matrix 
calculated excitation curves are shown in Figure~\ref{fig:Result}. Also presented is 
a summary of previously reported parameters.}
\begin{ruledtabular}
\begin{tabular}{c|cccc|clcccc}
&\multicolumn{4}{l}{This work}\vline&\multicolumn{6}{l}{Previous Results}\\
 Resonance&$E_{CM}$&$J^{\pi}$&$\Gamma_{p}$&$\Gamma_{\alpha}$
&$E_{CM}$&$J^{\pi}$&$\Gamma_{p}$&$\Gamma_{\alpha}$
&$\Gamma$&Ref.\\ 
&(MeV)&&(keV)&(keV)&(MeV)&&(keV)&(keV)&(keV)&\\ \hline
A\footnote{Parameters of state A are constrained to those shown here to provide normalisation. Hence, no parameters or error bars have been extracted in the present work.}&0.665&$\frac{3}{2}^{+}$&15.2&23.8&0.6647(16)&$\frac{3}{2}^{+}$&15.2(1)&23.8(12)&&\cite{Bardayan2001}\\
B&0.759(20)&$\frac{3}{2}^{+}$&1.6(5)&2.4(6)&
0.827(6)&$\frac{3}{2}^{+}$&0.35(35)&6.0(52)&&\cite{Bardayan2001}\\
&&&&
&0.793(31)&$\frac{3}{2}^{(+)}$&&&35(12)&\cite{Dalouzy2009}\\
C&1.096(11)&$\frac{5}{2}^{+}$&3(1)&54(12)
&1.089(9)&$\frac{5}{2}^{+}$&1.25(125)&0.24(24)&&\cite{Bardayan2004}\\
&&&&
&1.092(30)&$\frac{5}{2}^{(-)}$&&&17(7)&\cite{Dalouzy2009}\\
&&&&
&1.089(3)&$\frac{5}{2}^{+}$&1(1)&1.5(10)&&\cite{Murphy2009}\\
D&1.160(34)&$\frac{3}{2}^{+}$&2.3(6)&1.9(6)
&1.197(11)&$\frac{3}{2}^{+}$&2(1)&43(15)&&\cite{Utku1998}\\
&&&&
&1.206(5)&$\frac{3}{2}^{(+)}$&&&21(10)&\cite{Dalouzy2009}\\
E&1.219(22)&$\frac{3}{2}^{-}$&21(3)&0.1(1)
&1.233(12)&$\frac{1}{2}^{-}$&27(10)&16(6)&&\cite{Utku1998}\\
&&&&
&1.233(18)&$\frac{3}{2}^{-}$&1(1)&3(3)&&\cite{Murphy2009}\\
F&1.335(6)&$\frac{3}{2}^{+}$&65(8)&26(4)
&1.347(5)&$\frac{3}{2}^{+}$&42(10)&5(2)&&\cite{Murphy2009}\\
G&1.455(38)&$\frac{1}{2}^{+}$&55(12)&347(92)
&1.452(39)&$\frac{1}{2}^{(+)}$&&&292(107)&\cite{Dalouzy2009}\\
H&1.571(13)&$\frac{5}{2}^{+}$&1.7(4)&12(3)
&1.564(10)&$\frac{5}{2}^{(-)}$&&&11(8)&\cite{Dalouzy2009}\\
&&&&
&1.573(8)&$\frac{1}{2}^{+}$&8($^{+8}_{-4}$)&34(13)&&\cite{Murphy2009}\\
\end{tabular}
\end{ruledtabular}
\end{table*}

\begin{table*}
\caption{\label{tab:Covar} Covariance matrix for all parameters allowed to vary in the fitting process}
\begin{ruledtabular}
\begin{tabular}{c@{\,}c@{\,}c@{\,}c@{\,}c@{\,}c@{\,}c@{\,}c@{\,}c@{\,}c@{\,}c@{\,}c@{\,}c@{\,}c@{\,}c@{\,}c@{\,}c@{\,}c@{\,}c@{\,}c@{\,}c@{\,}|@{\,}c@{\,}c@{\,}}
\multicolumn{3}{l}{B}&\multicolumn{3}{l}{C}&\multicolumn{3}{l}{D}&\multicolumn{3}{l}{E}&\multicolumn{3}{l}{F}&\multicolumn{3}{l}{G}&\multicolumn{3}{l}{H}&&\\
$E_{CM}$&$\Gamma_{p}$&$\Gamma_{\alpha}$&$E_{CM}$&$\Gamma_{p}$&$\Gamma_{\alpha}$
&$E_{CM}$&$\Gamma_{p}$&$\Gamma_{\alpha}$&$E_{CM}$&$\Gamma_{p}$&$\Gamma_{\alpha}$&$E_{CM}$
&$\Gamma_{p}$&$\Gamma_{\alpha}$&$E_{CM}$&$\Gamma_{p}$&$\Gamma_{\alpha}$&$E_{CM}$
&$\Gamma_{p}$&$\Gamma_{\alpha}$&Par.&\\ \hline
1&-0.51&0.99&0&0&0&0&0&0&0&0&0&0&0.02&0.02&-0.01&0.01&0&0&0&0&$E_{CM}$&B\\
&1&-0.51&0.01&0.03&0&0&0&0&-0.01&0.07&-0.02&0&-0.08&0.01&0&-0.03&0.04&0&0&0&$\Gamma_{p}$&\\
&&1&0&0.01&0.01&0&0&0.01&0&0.01&0.01&-0.02&0.03&0.03&-0.02&0.01&0.02&0&0&0&$\Gamma_{\alpha}$&\\
&&&1&0.14&0.03&0.05&0.02&0.05&-0.02&0.01&0.50&-0.01&0.01&0.04&0.12&-0.14&0.02&-0.01&-0.02&-0.01&$E_{CM}$&C\\
&&&&1&-0.27&0.08&-0.02&0.07&-0.09&-0.05&-0.11&0.02&0.01&0.07&-0.27&0.21&0.25&0.09&0.05&0.11&$\Gamma_{p}$&\\
&&&&&1&0.02&0.13&0.07&-0.01&-0.02&-0.05&-0.02&0.06&0.07&0.10&0.21&-0.25&0.12&0.11&0.22&$\Gamma_{\alpha}$&\\
&&&&&&1&-0.78&0.99&-0.02&0.02&0.10&-0.03&0.08&0.08&-0.07&0.05&0.01&0&0&0&$E_{CM}$&D\\
&&&&&&&1&-0.71&-0.04&0.15&0.07&-0.01&-0.04&0.02&0.02&-0.05&0.05&0&0&-0.01&$\Gamma_{p}$&\\
&&&&&&&&1&-0.01&0.02&0.17&-0.08&0.13&0.12&-0.07&0.03&0.03&0&-0.01&-0.01&$\Gamma_{\alpha}$&\\
&&&&&&&&&1&-0.11&0.11&-0.02&0.04&0.04&-0.13&0.03&0.10&0&0.01&0&$E_{CM}$&E\\
&&&&&&&&&&1&-0.29&0.01&-0.42&0.07&-0.13&0.03&0.06&0&0&-0.01&$\Gamma_{p}$&\\
&&&&&&&&&&&1&-0.08&0.24&0.18&0.02&-0.46&0.27&-0.13&-0.20&-0.34&$\Gamma_{\alpha}$&\\
&&&&&&&&&&&&1&-0.50&0.17&0.04&0.08&-0.22&0&0&0&$E_{CM}$&F\\
&&&&&&&&&&&&&1&-0.02&-0.23&0.18&-0.08&-0.01&0.02&0.02&$\Gamma_{p}$&\\
&&&&&&&&&&&&&&1&-0.56&0.33&-0.08&-0.01&0.03&0.01&$\Gamma_{\alpha}$&\\
&&&&&&&&&&&&&&&1&-0.33&0&-0.33&0.22&0.19&$E_{CM}$&G\\
&&&&&&&&&&&&&&&&1&-0.69&-0.19&0.53&0.60&$\Gamma_{p}$&\\
&&&&&&&&&&&&&&&&&1&0.03&-0.41&-0.51&$\Gamma_{\alpha}$&\\
&&&&&&&&&&&&&&&&&&1&-0.63&-0.39&$E_{CM}$&H\\
&&&&&&&&&&&&&&&&&&&1&0.93&$\Gamma_{p}$&\\
&&&&&&&&&&&&&&&&&&&&1&$\Gamma_{\alpha}$&\\
\end{tabular}
\end{ruledtabular}
\end{table*}

In addition to resonant structures, the presence of carbon in the target allows a possible contribution from
fusion-evaporation reactions. To investigate this, data were taken with the CH$_{2}$ target replaced with a thick natural carbon target.
The resulting yield of both protons and alpha particles was found to be small and insufficient to make a significant 
contribution to any of the proposed resonant structures. This conclusion was supported by the 
results of fusion-evaporation event rate estimates made using the LISE computer code~\cite{Tarasov2003}.

To extract best estimates of the parameters of the states forming these 7 resonances, the data have been compared to
$R$-matrix calculations of the excitation functions under various assumptions for the energies and partial 
widths for each state. The $R$-matrix calculations~\cite{lane1958}
 were performed with the 
multichannel {\tt DREAM} code~\cite{Descouvemont2010},
 with a channel radius of~5~fm, and an energy dependent 
energy resolution (values determined by the Monte Carlo studies discussed earlier). 
Events detected were spread over a range of angles, from $\sim$172$^{\circ}$ to 180$^{\circ}$ 
in the center of mass. The value used for the calculations was, therefore, taken to be the average 
angle of all detector pixels, $\sim176^{\circ}$. It was found that there was no significant sensitivity 
in $\chi^{2}$ to alternative angle choices in the range $\sim$172$^{\circ}$ to 180$^{\circ}$.
The parameters resulting in a minimum in the reduced $\chi^{2}$ for the simultaneous fit to both data 
sets was searched for. In the region of minima, likelihoods were also calculated and it was observed that the maximum 
likelihood coincided with the minimum  $\chi^{2}$. The process has been repeated under alternative assumptions 
of angular momentum transfer and spin of the states, and for alternative possibilities for the signs of the interference 
between states of identical spin-parity. Under the condition that the widths and energies for all states 
(except the well-known 3/2$^{+}$ state at 7076~keV) could vary freely, the parameter set resulting in the overall 
smallest reduced $\chi^{2}$ is shown in Table~\ref{tab:Results}. Additionally, previous measurements of the states observed here are noted in the table. It can be seen that the assigned parameters vary significantly between many of the measurements. 

As noted in~\cite{Descouvemont2004}, the parameter uncertainty estimates calculated by the {\tt DREAM} code are underestimated 
when the minimum  $\bar{\chi}^{2}$ is greater than 1; the suggested procedure for correction to the uncertainties has been applied, 
{\em i.e.} the error bars on data points are increased to give the best fit line a $\bar{\chi}^{2}=1$,  and then
the parameter error estimation routines of the {\tt DREAM} code are reimplemented, generating revised estimates of the errors.
We have also explored the correlation between extracted parameters, and, unsurprisingly,  find several of the parameters 
to be strongly correlated, meaning the corresponding uncertainties are not independent. 
For completeness, Table~\ref{tab:Covar} lists the covariance matrix with large values 
highlighting parameters which are highly correlated.

The feature labelled B is likely to be the resonance previously observed by Bardayan {\em et al.}~\cite{Bardayan2001},
and by Dalouzy {\em et al.}~\cite{Dalouzy2009} (see Table~\ref{tab:Results} for observed energies and widths). 
The energy reported here is somewhat lower than previously observed, though it is poorly
constrained as is illustrated by the large covariance between the energy and relative widths of this state. 

Feature C has previously been observed several times~\cite{Utku1998, Bardayan2004,Murphy2009, Dalouzy2009},
but while there is reasonable agreement in energy and proton width, the present fit suggests a 
significantly broader alpha width than the earlier works, even given their disagreement. 
Re-analysis of the data in Murphy {\em et al.}~\cite{Murphy2009} has revealed an ambiguity such 
that a significantly broader alpha partial width than reported would also adequately describe the structure. 
Based on two previous observations of a state with likely $J^{\pi}=3/2^{+}$ approximately 100~keV higher 
in energy~\cite{Dalouzy2009,Utku1998}, we have included resonance D to our $R$-matrix calculations. 
We find a proton partial width for this state in agreement with Utku~{\em et al.}~\cite{Utku1998}, but a significantly 
narrower alpha width, and consequently a total width which is less than was found by Dalouzy~{\em et al.}~\cite{Dalouzy2009}. 
It seems likely therefore that features C and D are poorly resolved in this and other measurements, with 
both states having proton widths of order 1-3~keV and the sum of their alpha widths being of order 40-50 keV
(assuming their spins are as stated). 
Table~\ref{tab:Covar} suggests that the correlation between the extracted parameters for resonances C and D 
is low in the present measurement. 

The strong feature at $E_{CM}$=1.2--1.4~MeV is well described by two previously observed 
states in $^{19}$Ne, labelled here as E and F, but only when additional strength underlying
these states is attributed to an additional broad state, labelled G, described shortly. 
States E and F match those previously observed at excitation energies of 7624 and 7748~keV~\cite{Murphy2009,Utku1998}, where state F must have the opposite sign of interference to state A to adequately fit the data. Consistent with~\cite{Murphy2009}, state F was given a negative sign of interference while other 3/2$^{+}$ states are assumed to have positive interference signs. 
Here the $J^{\pi}=3/2^{-}$ assignment is favored for resonance E, as in Murphy {\em et al.}~\cite{Murphy2009}, 
although the proton and alpha widths are closer to those seen by Utku {\em et al.}~\cite{Utku1998} where an 
assignment of $J^{\pi}=1/2^{-}$ was proposed.

\begin{figure}[htb]
\includegraphics[scale=0.48]{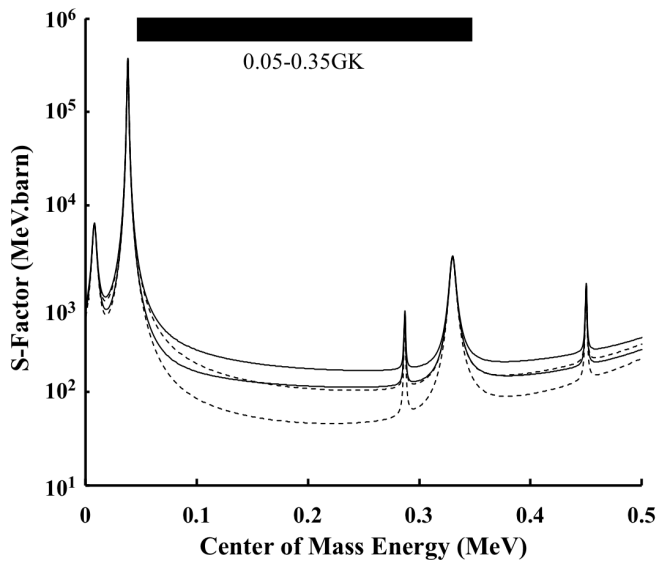}
\caption{\label{fig:s-factor} The Astrophysical $S$-factor for the $^{18}$F(p,$\alpha$)$^{15}$O reaction. 
The dashed lines are the astrophysical $S$-factor arising from interference combinations giving the highest and lowest values in the astrophysical region based on the parameters recommended by Iliadis {\em et al.}~\cite{Iliadis2010}. 
The addition of the predicted subthreshold $J^{\pi}=1/2^{+}$ state~\cite{Dufour2007} significantly 
enhances the $S$-factor, as shown by the solid lines.}
\end{figure}

A further clear feature, H,  is seen at about 1.571~MeV, and is here most well reproduced by a $J^{\pi}=5/2^{+}$ state. 
In the work of Murphy {\em et al.}~\cite{Murphy2009} a resonance was seen 
at close to the same energy, but a 1/2$^{+}$ assignment was found to best reproduce the data, although this was tentative due to the 
proximity of the state to the upper energy limit of that experiment. 
Dalouzy {\em et al.}~\cite{Dalouzy2009} also observed a state close to this energy, of 
somewhat smaller width, and made an unambiguous assignment of 5/2 for the spin based on a 
parity independent angular distribution measurement: the parity was inferred on the basis of the 
lower centrifugal barrier for protons in the reaction being studied.  In the present work,
attempts to fit the data with a $J^{\pi}=5/2^{-}$ assignment are poor, strongly favoring the 
positive parity assignment. 

As stated, a broad resonance, denoted G, has been included in the present work. Its contribution 
to the differential cross section of both reactions studied is illustrated by the dashed lines in Figure~\ref{fig:Result}.
Without the inclusion of this state, the best fit of the $R$-matrix calculations to the data is significantly worsened, 
especially in the higher energy region of the $^{18}$F(p,$\alpha$)$^{15}$O data. The overall reduced $\chi^{2}$ value for the
simultaneous fit to the entire data set changes from 1.633 to 2.483, and, considering only the data
between 1.0 and 1.8 MeV, the reduced $\chi^{2}$ changes from 1.306 to 2.486. Furthermore,
the deduced parameters for resonances C, D, E, F and H depart significantly further from literature values. 
With the inclusion of resonance G, our best fit corresponds to a $J^{\pi}=1/2^{+}$ state at an excitation
energy of 7870$\pm$40~keV with a proton partial width of 55$\pm$12~keV and an alpha partial width of 347$\pm$92~keV. 
The extracted parameters show quite strong correlations with those extracted for resonances F and H. 
The data are consistent with the presence of the state predicted by Dufour and Descouvemont~\cite{Dufour2007} and 
observed by Dalouzy~{\em et al.}~\cite{Dalouzy2009}. 

The apparent existence of the broad state G supports the prediction of an additional broad 1/2$^{+}$ state below threshold. 
The impact of such a state is illustrated in Figure~\ref{fig:s-factor}, where the astrophysical 
$S$-Factor most recently recommended by 
Iliadis {\em et al.}~\cite{Iliadis2010} is shown (dashed lines), together with same $S$-factor modified by the inclusion of the proposed 
subthreshold 1/2$^{+}$ state with parameters as suggested by~\cite{Dufour2007} (solid lines). The two curves shown in 
each case correspond to the highest and lowest $S$-factors allowed due to the uncertainty in the 
interference between the 38 and 665~keV $J^{\pi}= 3/2^{+}$ states. Within the Gamow window, the astrophysical $S$-factor
is increased, is more tightly constrained, and the possibility of strongly destructive interference, as
highlighted by de~S\'er\'eville {\em et al.}~\cite{Sereville2005}  is removed.
The lowest energy measurement to date~\cite{Beer2011} was at 250~keV, with an $S$-factor of 105$^{+118}_{-60}$~MeV~b, 
in good agreement with this result. 

\section{\label{sec:Conc}Conclusion}
\noindent
New data have been obtained in the study of the astrophysically important $^{18}$F(p,$\alpha$)$^{15}$O reaction,
relevant to gamma-ray production in novae. An $R$-matrix analysis has been performed to deduce the
parameters of states causing resonant structures in measured excitation functions.
A well-known state was clearly identified at $E_{CM}=0.665~MeV$ and constraints were 
placed on the parameters of 7 more resonances, including those populated by
low angular momentum transfers which are of most astrophysical importance.  
The results are consistent with a recent prediction and measurement of a broad
1/2$^{+}$ state at an excitation of~$\sim$7860~keV in $^{19}$Ne. The existence of this 
state supports the Generator Coordinate Method prediction of another broad state at sub-threshold energies, 
contributing significantly at novae temperatures. 
The inclusion of this resonant contribution is independent of the unknown strengths arising from 
several $J^{\pi}= 3/2^{+}$ states where the interference between the states generates a large uncertainty. 
Hence, the reaction rate at astrophysical temperatures is further constrained to
relatively high values, resulting in a lower abundance of $^{18}$F in nova ejecta 
and a consequent reduced detectability distance for these events. 

\begin{acknowledgments}
\noindent
The authors would like to thank the beam development and SPIRAL operations group at the GANIL facility. UK personnel were supported by the Science and Technologies Facilities Council (UK). The University of Edinburgh is a charitable body, registered in Scotland, with the registration number SC005336.
\end{acknowledgments}

\bibliography{18F}

\end{document}